\documentclass[a4paper,12pt]{article}
\usepackage{amsmath}
\usepackage{accents}
\usepackage{bm}
\usepackage{dsfont}
\usepackage[english]{babel}
\usepackage{graphicx}
\usepackage[usenames]{color}
\usepackage{colortbl}
\usepackage{mathtools}
\usepackage{commath}
\usepackage[font=footnotesize,figurewithin=none]{caption}
\usepackage[caption=false]{subfig}
\usepackage[toc,page,title]{appendix}

\begin{document}

\centerline {\LARGE{Preparation of two-qubit entangled states}}
\centerline {\LARGE{on a spin-1/2 Ising-Heisenberg diamond spin cluster}}
\centerline {\LARGE{by controlling the measurement}}
\medskip
\centerline {A. R. Kuzmak}
\centerline {\small \it E-Mail: andrijkuzmak@gmail.com}
\medskip
\centerline {\small \it Department for Theoretical Physics, Ivan Franko National University of Lviv,}
\medskip
\centerline {\small \it 12 Drahomanov St., Lviv, UA-79005, Ukraine}

\date{\today}
\begin{abstract}
The preparation of entangled quantum states is an inherent and indispensable step for the implementation of many quantum information algorithms. Depending on the physical system, there are different ways to control and measure them,
which allow one to achieve the predefined quantum states. The diamond spin cluster is the system that can be applied for this purpose. Moreover, such a system appears in chemical compounds such as the natural mineral azurite,
where the ${\rm Cu^{2+}}$ are arranged in a spin-$1/2$ diamond chain. Herein, we propose the method of preparation of pure entangled states on the Ising-Heisenberg spin-$1/2$ diamond cluster. We suppose that the cluster consists
of two central spins which are described by an anisotropic Heisenberg model and interact with the side spins via Ising interaction. Controlling the measurement direction of the side (central) spins allows us to achieve
predefined pure quantum states of the central (side) spins. We show that this directly affects the entanglement and fidelity of the prepared states. For example, we obtain conditions and fidelities for preparations of the Bell states.
\end{abstract}

\section{Introduction \label{sec1}}

The preparation of entangled states plays a crucial role in the implementation of quantum information algorithms \cite{desurvire2009}. Entangled states are an integral part of
quantum cryptography \cite{Ekert1991}, super-dense coding \cite{Bennett1992}, quantum teleportation \cite{TELEPORT,Zeilinger1997},
quantum calculations \cite{cerf1998,pittman2001,gasparoni2004}, optimization of quantum calculations \cite{Giovannetti20031,Giovannetti20032,Batle2005,Borras2006}, etc.
All these schemes require specific physical systems that can be easily controlled and measured. There are following systems that are used for this purpose: polarized photons \cite{ASPECT,Zeilinger1997,gasparoni2004,englert2001},
nuclear and electronic spins of atoms \cite{quantcomp,qdots1,phosphorus3,phosphorus1,kuzmak2020}, superconducting qubits \cite{supcond1,supcond2,supcond3,supcond4},
trapped ions \cite{SchrodCat1,EQSSTI,SchrodCat2,ETDIITIQSHI,QSDEGHTI}, ultracold atoms \cite{opticallattice1,opticallattice18,opticallattice5}, etc. In the last years, the preparation of entangled states
and their application to the algorithms of quantum information is also widely studied on quantum computers \cite{wang2018s,mooney2019,kuzmak20201,arute2019,kuzmak20202,kuzmak2021,gnatenko2021,gnatenko20212}.

One of the system which can be applied for the quantum information is a diamond spin cluster formed by four spins. Many compounds contain arranged diamond spin clusters in chains. For instance, there are the following copper-based compounds:
${\rm Ca_3Cu_3(PO_4)_4}$, ${\rm Sr_3Cu_3(PO_4)_4}$ \cite{drillon1988,drillon1993} ${\rm Bi_4Cu_3V_2O_{14}}$ \cite{sakurai2002},${\rm Cu_3(CO_3)_2(OH)_2}$ (also called  the natural mineral azurite) \cite{kikuchi2005}.
The ${\rm Cu^{2+}}$ ions in the natural mineral azurite form a spin-1/2 diamond chain. Recently quantum properties such as the entanglement of these systems have been actively examined.
In papers \cite{bose2005,tribedi2006,ananikian2006,ananikian2012,chakhmakhchyan2012,rojas2012,rojas2014,torrico2016,rojas2017,Zheng2018,Cavalho2019,Ghannadan2022} the bipartite entanglement between spins in the diamond spin cluster which are in thermodynamic equilibrium was studied.
Bose and Tribedi were provided the first calculations of thermal entanglement in the diamond spin cluster \cite{bose2005,tribedi2006}. Thermal entanglement of a spin-1/2 Ising-Heisenberg symmetrical diamond chain was studied for the first time by Nerses Ananikian et all  \cite{ananikian2006}. They calculated the behaviour of entanglement as a function of the system parameters. They also showed that for a dominant Heisenberg-type interaction the system’s ground state is maximally entangled, but on increasing the temperature pure quantum correlations disappear.
For other types of interaction between spins such as $XXZ$-Heisenberg \cite{rojas2012}, $XYZ$-Heisenberg \cite{rojas2014}, Ising-$XYZ$ distorted diamond \cite{rojas2017}, etc. the behaviour of thermal entanglement was also investigated.
Recently, thermal entanglement, local quantum uncertainty, and quantum coherence in a four-qubit square chain described by the Heisenberg $XXZ$ Hamiltonian was exactly examined \cite{Benabdallah2022}.
The authors studied the influences of the Hamiltonian parameters on these criteria and fidelity of teleportation.
In our previous paper \cite{kuzmak2023}, for the first time, we studied the bipartite entanglement of the  Ising-Heisenberg diamond spin-1/2 cluster in evolution.

Control of quantum systems plays an important role in the preparation of states. Different types of systems are controlled in a specific way. The evolution of spin systems
is controlled by the values of interaction between spins and external magnetic fields. The predefined states are achieved on the system by the measurenents at a certain moment of time.
The technique which allows one to control and measure such systems is called the spin resonance technique \cite{srtech,Vandersypen2004}.
Implementation of quantum states on different spin systems was widely studied in papers \cite{Nichol2017,Harvey-Collard2018,Nagy2019,Kuzmak2014,Kuzmak2018,Sahling2015,twosqg2,twosqg3,twosqg4,twosqg5,twosqg6,Zu2014,kuzmak2020}.

In this paper we consider the preparation of two-qubit pure entangled states on the central (side) spins of the diamond cluster by controlling the measurement directions of side (central) spins.
The diamond spin cluster consists of two central spins described by anisotropic Heisenberg Hamiltonian which interact with two side spins via the Ising model (Sec.~\ref{modelsec}). In Secs.~\ref{entastatesab}, \ref{statesspin12}, the achievement of entangled states on central and side spins is studied. The conditions for achieving maximally entangled states are obtained. For example, the preparation of Bell states on the central and side spins is considered.

\section{Model of a diamond spin cluster \label{modelsec}}

\begin{figure}[!!h]
\centerline{\includegraphics[scale=0.60, angle=0.0, clip]{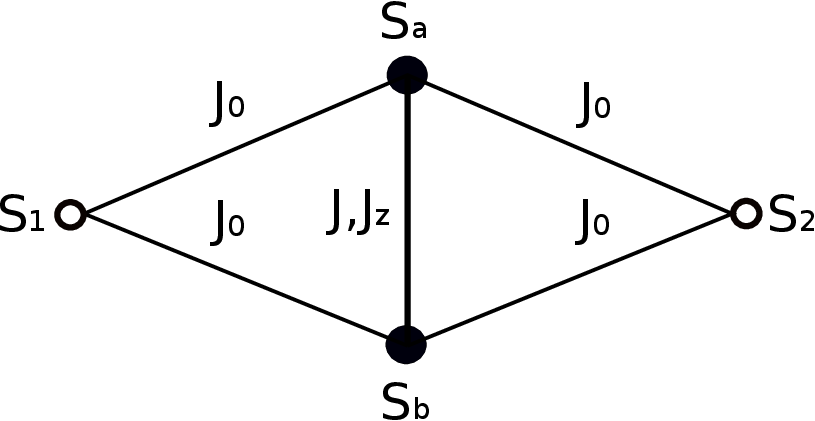}}
\caption{The structure of a diamond spin cluster consists of two central spins $S_a$, $S_b$ described by anisotropic Heisenberg interaction and which interact with two side spins $S_1$, $S_1$ via Ising interaction.}
\label{model}
\end{figure}

We consider the diamond spin cluster that consists of two central $S_a$, $S_b$ spin-$1/2$ described by anisotropic Heisenberg Hamiltonian and two side spin-$1/2$ $S_1$, $S_2$ (Fig.~\ref{model}).
Interaction of the central spins with the side spins is defined by the Ising model.
The Hamiltonian of the whole system is expressed by three terms that mutually commute
\begin{eqnarray}
H=H_{ab}+H_{12}+H_{int},
\label{hamiltonian}
\end{eqnarray}
where
\begin{eqnarray}
&&H_{ab}=J\left(S_{a}^xS_{b}^x+S_{a}^yS_{b}^y\right)+J_zS_{a}^zS_{b}^z+h'\left(S_{a}^z+S_{b}^z\right),\label{hamab}\\
&&H_{12}=h\left(S_{1}^z+S_{2}^z\right),\label{ham12}\\
&&H_{int}=J_0\left(S_{a}^z+S_{b}^z\right)\left(S_{1}^z+S_{2}^z\right)\label{hamint}.
\end{eqnarray}
Here ${\bf S}_{\alpha}=1/2\left(S_{\alpha}^x{\bf i}+S_{\alpha}^y{\bf j}+S_{\alpha}^z{\bf k}\right)$ is the operator of $\alpha$-th spin ($\alpha=a,b,1,2$), $J$ and $J_z$ are the coupling constants between $a$ and $b$ spins,
$J_0$ is a coupling constant which defines the interaction between $S_a$, $S_b$ and $S_1$, $S_2$ pairs of spins, $h'$, $h$ are the values which describe the interaction between spins and an external magnetic field.
We use the system of units, where the Planck constant is $\hbar =1$. This means that the energy is measured in units of the frequency. The Hamiltonians $H_{ab}$ (\ref{hamab}), $H_{12}$ (\ref{ham12})
describe the subsystems of $S_a$, $S_b$ and $S_1$, $S_2$ spins, respectively, and the Hamiltonian $H_{int}$ (\ref{hamint}) describes the interaction between those subsystems.
Since these Hamiltonians mutually commute
\begin{eqnarray}
\left[H_{ab},H_{12}\right]=\left[H_{ab},H_{int}\right]=\left[H_{12},H_{int}\right]=0,\nonumber
\end{eqnarray}
we can easily find eigenvalues and eigenstates of the system (see Appendix~\ref{appegen}).

The evolution of diamond spin cluster determined by Hamiltonian (\ref{hamiltonian}) having started from the initial state $\vert\psi_I\rangle$ can be expressed as follows
\begin{eqnarray}
\vert\psi(t)\rangle=e^{-iHt}\vert\psi_I\rangle=e^{-iHt}\sum_n C_n\vert\psi_n\rangle=\sum_n C_ne^{-iE_nt}\vert\psi_n\rangle,
\label{evolution}
\end{eqnarray}
where $\vert\psi_n\rangle$ and $E_n$ is a set of eigenstates and eigenvalues given by expression (\ref{eigenvaleigenstate}), $C_n$ are the complex parameters that determine the initial state.
Controlling the initial state, the values of the external magnetic field, and the time of evolution, allows us to achieve the predefined final state. Choosing the measurement direction of one subsystem allows us to fix the final pure state of another subsystem. Moreover we can achieve the predefined pure entangled states. Let us consider the preparation of these states on the $S_a$, $S_b$ spins by controlling the measurement direction of the $S_1$, $S_2$ spins, and vice versa.

The copper-based compounds mentioned above form a spin chain. We consider the evolution of a separate diamond spin cluster. In the case of chain the interaction between the Heisenberg spins is provided by the
Ising spins. In other words, the dimers of the chain interact between themselves via Ising spins which leads to the fact that all spins of the chain affect the state of the selected diamond cluster.
However, when Ising spins are in the eigenstates, the evolution of all dimer spins in the chain is independent. This fact allows one to consider the evolution of a single dimer in a diamond spin cluster (Subsec.~\ref{entastatesab1}).

\section{Preparation of entangled states on the $S_a$, $S_b$ spins \label{entastatesab}}

In this section, we examine the preparation of quantum states on the $S_a$ and $S_b$ spins. We consider two cases of the evolution of the whole system: 1. the side $S_1$, $S_2$ spins do not evolve; 2. the side $S_1$, $S_2$ spins evolve. Depending on the measurement of the side spins, we obtain the conditions for the preparation of entangled pure states on the $S_a$, $S_b$ spins. We calculate the entanglement of these states. For this purpose, we use the Wootters definition of concurrence \cite{wootters1997,wootters1998}
\begin{eqnarray}
C(\vert\psi\rangle)=2\vert ad-bc\vert,
\label{wootterspure}
\end{eqnarray}
where $a$, $b$, $c$ and $d$ are complex parameters which define the state
\begin{eqnarray}
\vert\psi\rangle =a\vert\uparrow\uparrow\rangle+b\vert\uparrow\downarrow\rangle+c\vert\downarrow\uparrow\rangle+d\vert\downarrow\downarrow\rangle.
\label{purestate}
\end{eqnarray}
and satisfy the normalization condition $\vert a\vert^2+\vert b\vert^2+\vert c\vert^2+\vert d\vert^2=1$. The concurrence takes values $C=\left[0,1\right]$. For the separated states, it equals $C=0$ and for the maximally entangled states, it takes the value $C=1$.

\subsection{Stationarity of the $S_1$, $S_2$ spins \label{entastatesab1}}

Let us consider the evolution of $S_{a}$, $S_{b}$ spins without including in the evolution of $S_1$, $S_2$ spins. For this purpose, we prepared the initial state of the system in the way that the spins $S_1$ and $S_2$ do not evolve. Due to the facts that Hamiltonians $H_{ab}$ (\ref{hamab}) and $H_{12}$ (\ref{ham12}) mutually commute, we take the eigenstate of Hamiltonian (\ref{ham12}) in the initial state. It does not matter which we take the eigenstate of the $S_1$, $S_2$ spins because this subsystem does not evolve. Thus the initial state of the whole system we take in the form
\begin{eqnarray}
\vert\psi_I\rangle=\vert\uparrow\uparrow\rangle_{12}\left(C_1\vert\uparrow\rangle_a+C_2\vert\downarrow\rangle_a\right)\left(C_3\vert\uparrow\rangle_b+C_4\vert\downarrow\rangle_b\right),
\label{initstaab}
\end{eqnarray}
where $C_1$, $C_2$, $C_3$ and $C_4$ are the complex parameters which define the initial state of $S_a$, $S_b$ spins and satisfy the following normalization conditions: $\vert C_1\vert^2+\vert C_2\vert^2=1$,
	$\vert C_3\vert^2+\vert C_4\vert^2=1$. The initial state can be decomposed by eigenstates (\ref{eigenvaleigenstate}) as follows
\begin{eqnarray}
&&\vert\psi_I\rangle=C_1C_3\vert\psi_1\rangle+C_1C_4\frac{1}{\sqrt{2}}\left(\vert\psi_2\rangle+\vert\psi_3\rangle\right)\nonumber\\
&&+C_2C_3\frac{1}{\sqrt{2}}\left(\vert\psi_2\rangle-\vert\psi_3\rangle\right)+C_2C_4\vert\psi_4\rangle.
\label{initstaabrew}
\end{eqnarray}
Based on the equation (\ref{evolution}), the evolution of the system takes the form
\begin{eqnarray}
&&\vert\psi(t)\rangle=e^{-iHt}\vert\psi_I\rangle\nonumber\\
&&=C_1C_3e^{-iE_1t}\vert\psi_1\rangle+C_1C_4\frac{1}{\sqrt{2}}\left(e^{-iE_2t}\vert\psi_2\rangle+e^{-iE_3t}\vert\psi_3\rangle\right)\nonumber\\
&&+C_2C_3\frac{1}{\sqrt{2}}\left(e^{-iE_2t}\vert\psi_2\rangle-e^{-iE_3t}\vert\psi_3\rangle\right)+C_2C_4e^{-iE_4t}\vert\psi_4\rangle\nonumber\\
&&=C_1C_3e^{-i(h+\frac{J_z}{4}+h'+J_0)t}\vert\psi_1\rangle+C_1C_4\frac{1}{\sqrt{2}}e^{-i(h-\frac{J_z}{4})t}\left(e^{-i\frac{Jt}{2}}\vert\psi_2\rangle+e^{i\frac{Jt}{2}}\vert\psi_3\rangle\right)\nonumber\\
&&+C_2C_3\frac{1}{\sqrt{2}}e^{-i(h-\frac{J_z}{4})t}\left(e^{-i\frac{Jt}{2}}\vert\psi_2\rangle-e^{i\frac{Jt}{2}}\vert\psi_3\rangle\right)+C_2C_4e^{-i(h+\frac{J_z}{4}-h'-J_0)t}\vert\psi_4\rangle.\nonumber\\
\label{evolution1}
\end{eqnarray}
In the basis $\vert\uparrow\uparrow\rangle_{ab}$, $\vert\uparrow\downarrow\rangle_{ab}$, $\vert\downarrow\uparrow\rangle_{ab}$ and  $\vert\downarrow\downarrow\rangle_{ab}$ this state can be represented as follows
\begin{eqnarray}
&&\vert\psi(t)\rangle=e^{-iht}\vert\uparrow\uparrow\rangle_{12}\left[C_1C_3e^{-i(\frac{J_z}{4}+h'+J_0)t}\vert\uparrow\uparrow\rangle_{ab}\right.\nonumber\\
&&\left.+e^{i\frac{J_z}{4}t}\left(C_1C_4\cos\left(\frac{Jt}{4}\right)-iC_2C_3\sin\left(\frac{Jt}{4}\right)\right)\vert\uparrow\downarrow\rangle_{ab}  \right. \nonumber\\
&&\left.+e^{i\frac{J_z}{4}t}\left(C_2C_3\cos\left(\frac{Jt}{4}\right)-iC_1C_4\sin\left(\frac{Jt}{4}\right)\right)\vert\downarrow\uparrow\rangle_{ab}  \right.\nonumber\\
&&\left. +C_2C_4e^{-i(\frac{J_z}{4}-h'-J_0)t}\vert\downarrow\downarrow\rangle_{ab}\right].
\label{evolution2}
\end{eqnarray}
The interaction with the side $S_1$, $S_2$ spins affects the central $S_a$, $S_b$ spins as an effective magnetic field of the value $J_0$. During the evolution, the states of the side and central spins remain separate. Thus the state of $S_a$, $S_b$ spins does not depend on the measurements of the $S_1$ and $S_2$ spins. Selection of the initial state, system parameters, and the period of evolution allows us to achieve different entangled states. Using definition (\ref{wootterspure}), we calculate the value of entanglement of this state
\begin{eqnarray}
&&C(\vert\psi(t)\rangle_{ab})\nonumber\\
&&=2\left\vert C_1C_2C_3C_4\left(1-e^{iJ_zt}\cos(Jt)\right)+\frac{i}{2}e^{iJ_zt}\left(C_1^2C_4^2+C_2^2C_3^2\right)\sin(Jt)\right\vert,\nonumber\\
\label{woottersabpure}
\end{eqnarray}
where $\vert\psi(t)\rangle_{ab}$ is the state of $S_a$, $S_b$ spins that is separated from the state of $S_1$, $S_2$ spins in equation (\ref{evolution2}). For example, let us obtain the conditions for the preparation of different entangled states.

\begin{figure}[!!h]
\includegraphics[scale=0.53, angle=0.0, clip]{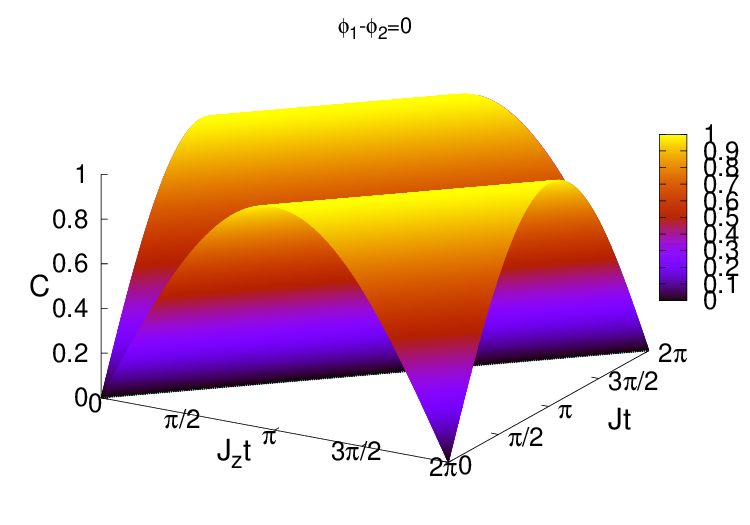}
\includegraphics[scale=0.53, angle=0.0, clip]{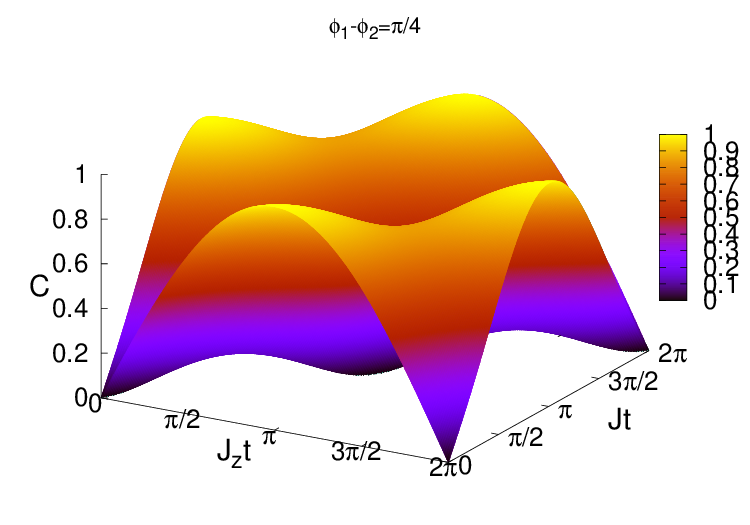}\\
\includegraphics[scale=0.53, angle=0.0, clip]{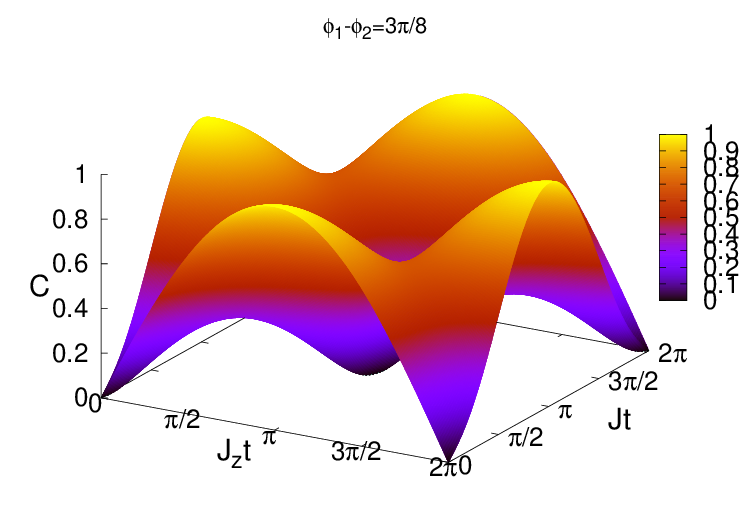}
\includegraphics[scale=0.53, angle=0.0, clip]{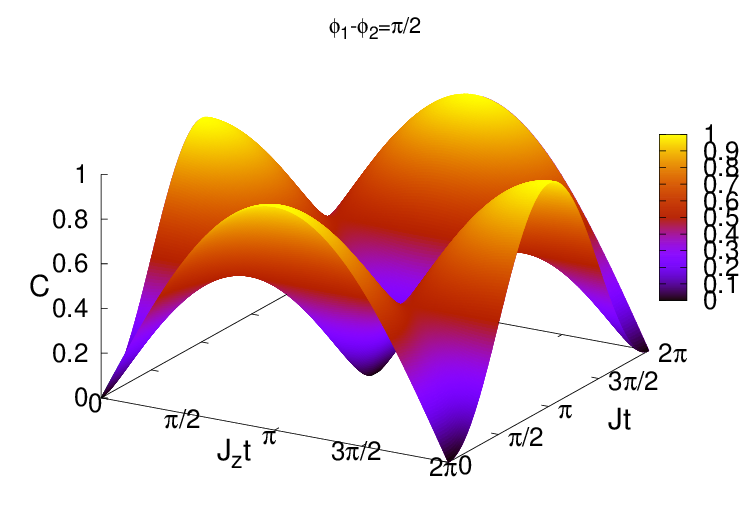}\\
\includegraphics[scale=0.53, angle=0.0, clip]{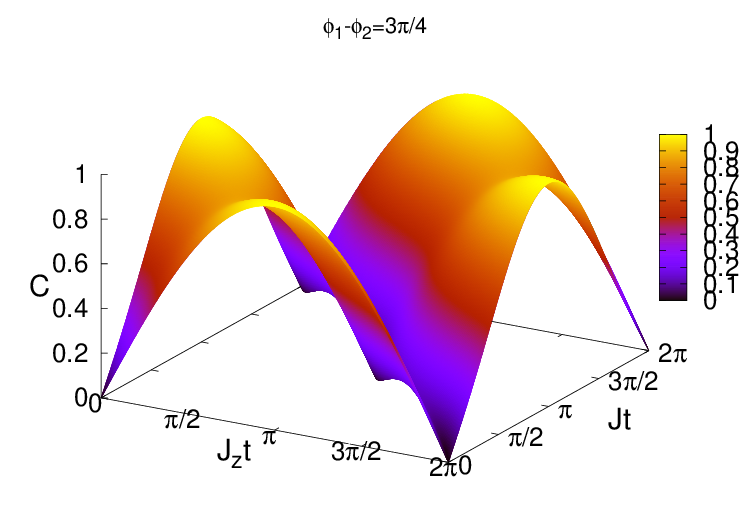}
\includegraphics[scale=0.53, angle=0.0, clip]{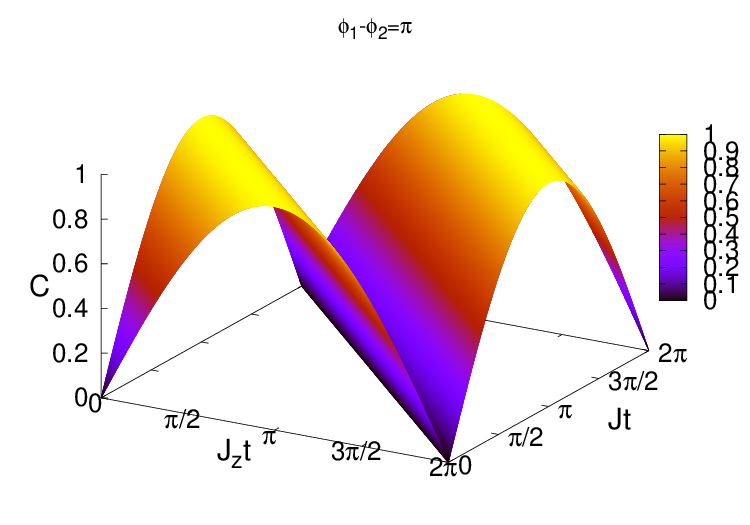}
\caption{Time dependences of concurrence (\ref{concabstat12}) on the interaction parameters between $S_a$ and $S_b$ spins for different values of $\phi_1-\phi_2$. Changing the value $\phi_1-\phi_2$ and ratio
between interaction couplings allows us to control the value of entanglement and time of reaching the entagled states.}
\label{entabdelta}
\end{figure}

Suppose that $C_1=C_4=1$ and $C_2=C_3=0$ than the initial state has the form
$\vert\psi_I\rangle=\vert\uparrow\uparrow\rangle_{12}\vert\uparrow\downarrow\rangle_{ab}$. Entangled state reached during the evolution of $S_a$, $S_b$ spins is the following
\begin{eqnarray}
\vert\psi(t)\rangle_{ab}=\cos\left(Jt/4\right)\vert\uparrow\downarrow\rangle-i\sin\left(Jt/4\right)\vert\uparrow\downarrow\rangle.\nonumber
\end{eqnarray}
From the equation (\ref{woottersabpure}) follows that the concurrence of this state has the form
\begin{eqnarray}
C(\vert\psi(t)\rangle_{ab})=\vert\sin(Jt)\vert.\nonumber
\end{eqnarray}
It takes the maximal values in the moments which satisfy the condition $Jt_n=\pi/2+\pi n$, where $n\in\mathds{Z}$.

Let us now project the initial state of $S_a$, $S_b$ spins in the plane $xy$. Then the parameters of initial state take the form $C_1=1/\sqrt{2}$, $C_2=e^{i\phi_1}/\sqrt{2}$, $C_3=1/\sqrt{2}$, $C_4=e^{i\phi_2}/\sqrt{2}$,
where $\phi_1$ and $\phi_2$ are the azimuthal angles of spherical coordinate system. For these parameters concurrence (\ref{woottersabpure}) reduces to the equation
\begin{eqnarray}
C(\vert\psi(t)\rangle_{ab})=\frac{1}{2}\left[\left(\cos(J_zt)-\cos(Jt)\right)^2+\left(\sin(J_zt)-\sin(Jt)\cos(\phi_1-\phi_2)\right)^2\right]^{1/2}.\nonumber\\
\label{concabstat12}
\end{eqnarray}
Controlling the parameters $J$, $J_z$, $\phi_1$, $\phi_2$, and time of evolution allows us to achieve entangled states. For instance, if we put $\phi_1-\phi_2=0$ and $(J_z-J)t=(2n+1)\pi$ or $\phi_1-\phi_2=\pi$ and $(J_z+J)t=(2n+1)\pi$
($n\in\mathds{Z}$), we obtain the maximally entangled states of $S_a$, $S_b$ spins ($C(\vert\psi(t)\rangle_{ab})=1$). Dependencies of concurrence (\ref{concabstat12}) for different values of $\phi_1-\phi_2$ are presented
in Fig.~\ref{entabdelta}.

\subsection{Dynamics of the $S_1$, $S_2$ spins \label{entastatesab2}}

Now we consider the case when the whole system evolves. For this purpose, we prepare the initial state as a projection of the spins in the positive direction of the $x$-axis. This state can be expressed as follows
\begin{eqnarray}
&&\vert\psi_I\rangle=\frac{1}{4}\left(\vert\uparrow\uparrow\rangle_{12}+\vert\uparrow\downarrow\rangle_{12}+\vert\downarrow\uparrow\rangle_{12}+\vert\downarrow\downarrow\rangle_{12}\right)\nonumber\\
&&\left(\vert\uparrow\uparrow\rangle_{ab}+\vert\uparrow\downarrow\rangle_{ab}+\vert\downarrow\uparrow\rangle_{ab}+\vert\downarrow\downarrow\rangle_{ab}\right).
\label{initstate12ab}
\end{eqnarray}
Taking into account equation (\ref{evolution}) the evolution of the system takes the form
\begin{eqnarray}
&&\vert\psi(t)\rangle=e^{-iHt}\vert\psi_I\rangle=\frac{1}{4}\left(e^{-iE_1t}\vert\psi_1\rangle+\sqrt{2}e^{-iE_2t}\vert\psi_2\rangle + e^{-iE_4t}\vert\psi_4\rangle \right.\nonumber\\
&&\left.+e^{-iE_5t}\vert\psi_5\rangle+\sqrt{2}e^{-iE_6t}\vert\psi_6\rangle+e^{-iE_8t}\vert\psi_8\rangle+ e^{-iE_9t}\vert\psi_9\rangle+\sqrt{2}e^{-iE_{10}t}\vert\psi_{10}\rangle\right.\nonumber\\
&&\left.+e^{-iE_{12}t}\vert\psi_{12}\rangle+e^{-iE_{13}t}\vert\psi_{13}\rangle+\sqrt{2}e^{-iE_{14}t}\vert\psi_{14}\rangle+e^{-iE_{16}t}\vert\psi_{16}\rangle\right),
\label{evolution12ab}
\end{eqnarray}
where the initial state is decomposed by the eigenstates of Hamiltonian (\ref{hamiltonian}) with eigenvalues $E_i$ (\ref{eigenvaleigenstate}). We express this state in the following form
\begin{eqnarray}
\vert\psi(t)\rangle=\frac{1}{2}\left(\vert\xi_1\rangle_{ab}\vert\uparrow\uparrow\rangle_{12}+ \vert\xi_2\rangle_{ab}\left(\vert\uparrow\downarrow\rangle_{12}+\vert\downarrow\uparrow\rangle_{12}\right) + \vert\xi_3\rangle_{ab}\vert\downarrow\downarrow\rangle_{12}\right),
\label{evolution12ab2}
\end{eqnarray}
where
{\small
\begin{eqnarray}
&&\vert\xi_1\rangle_{ab}=\frac{1}{2}\left[e^{-i(\frac{J_z}{4}+J_0+h+h')t}\vert\uparrow\uparrow\rangle_{ab}+e^{-i(\frac{J}{2}-\frac{J_z}{4}+h)t}(\vert\uparrow\downarrow\rangle_{ab}+\vert\downarrow\uparrow\rangle_{ab})+e^{-i(\frac{J_z}{4}-J_0+h-h')t}\vert\downarrow\downarrow\rangle_{ab}     \right],\nonumber\\
&&\vert\xi_2\rangle_{ab}=\frac{1}{2}\left[e^{-i(\frac{J_z}{4}+h')t}\vert\uparrow\uparrow\rangle_{ab}+e^{-i(\frac{J}{2}-\frac{J_z}{4})t}(\vert\uparrow\downarrow\rangle_{ab}+\vert\downarrow\uparrow\rangle_{ab})+e^{-i(\frac{J_z}{4}-h')t}\vert\downarrow\downarrow\rangle_{ab}     \right],\nonumber\\
&&\vert\xi_3\rangle_{ab}=\frac{1}{2}\left[e^{-i(\frac{J_z}{4}-J_0-h+h')t}\vert\uparrow\uparrow\rangle_{ab}+e^{-i(\frac{J}{2}-\frac{J_z}{4}-h)t}(\vert\uparrow\downarrow\rangle_{ab}+\vert\downarrow\uparrow\rangle_{ab})+e^{-i(\frac{J_z}{4}+J_0-h-h')t}\vert\downarrow\downarrow\rangle_{ab}     \right].\nonumber\\
\label{purestateofabspins}
\end{eqnarray}}
Measuring the $S_1$, $S_2$ spins on the $z$-axis, we obtain the central $S_a$, $S_b$ spins in the states defined by expression (\ref{purestateofabspins}).
The fidelity of the state $\vert\psi_c\rangle$ is determined in the following way
\begin{eqnarray}
F=\vert\langle\psi(t)\vert\psi_c\rangle\vert^2.
\label{fidelity}
\end{eqnarray}
In table~\ref{tab1}, we present the set of results presented with corresponding fidelities.
\begin{table}[ht] 
\centering 
  \begin{tabular}{ | c| c | c |  }
    \hline
State of $S_1$, $S_2$ spins & State of $S_a$, $S_b$ spins        & $F$                 \\ \hline
$\vert\uparrow\uparrow\rangle_{12}$   & $\vert\xi_1\rangle_{ab}$ &  1/4          \\ \hline
$\vert\uparrow\downarrow\rangle_{12}$   & $\vert\xi_2\rangle_{ab}$ &  1/4             \\ \hline
$\vert\downarrow\uparrow\rangle_{12}$   & $\vert\xi_2\rangle_{ab}$ &  1/4            \\ \hline
$\vert\downarrow\downarrow\rangle_{12}$   & $\vert\xi_3\rangle_{ab}$ &  1/4            \\ \hline
  \end{tabular}
  \caption{Results of the measurement of $S_1$, $S_2$ spins in the state (\ref{evolution12ab2}) at the moment $t$.}
  \label{tab1} 
  \end{table}
Using the Wootters definition (\ref{wootterspure}), we calculate the value of entanglement of these states. For all three states (\ref{purestateofabspins}) it takes the form
\begin{eqnarray}
C=\left\vert\sin\left(\frac{J_z-J}{2}t\right)\right\vert.
\label{valentsasb}
\end{eqnarray}
The maximally entangled states are achieved at the moment $t=(2n+1)\pi/(J_z-J)$.

In addition to the period of evolution and the parameters included in the Hamiltonian, the direction in which the $S_1$ and $S_2$ spins are measured affects the form of achieved states. This fact is easy to show when the state of side spins is rewritten in the basis defined by some direction. Let us assume that this direction is defined by the spherical angles $\theta$ and $\phi$. Then states of spin-$1/2$ projected in this direction have the form
\begin{eqnarray}
\vert +\rangle=\cos\left(\frac{\theta}{2}\right)\vert\uparrow\rangle+\sin\left(\frac{\theta}{2}\right)e^{i\phi}\vert\downarrow\rangle,\quad \vert -\rangle=-\sin\left(\frac{\theta}{2}\right)e^{-i\phi}\vert\uparrow\rangle+\cos\left(\frac{\theta}{2}\right)\vert\downarrow\rangle,\nonumber\\
\label{possnegstates}
\end{eqnarray}
where $\vert +\rangle$, and $\vert -\rangle$ are the states which correspond to the positive and negative projections, respectively. The relations between bases of the $S_1$, $S_2$ spins are presented in Appendix~\ref{appproject}.

Substituting expressions (\ref{newbasisstate}) in state (\ref{evolution12ab2}), we rewrite it in the form
\begin{eqnarray}
\vert\psi(t)\rangle=\frac{1}{4}\left(\frac{1}{A_1}\vert\psi_1\rangle_{ab}\vert ++\rangle_{12}+ \frac{1}{A_2}\vert\psi_2\rangle_{ab}\left(\vert +-\rangle_{12}+\vert -+\rangle_{12}\right) + \frac{1}{A_3}\vert\psi_3\rangle_{ab}\vert --\rangle_{12}\right).
\label{evolution12ab3}
\end{eqnarray}
Measuring the $S_1$, $S_2$ spins on the $\vert +\rangle$, $\vert -\rangle$ basis, the $S_a$, $S_b$ spins are reduced to one of the following three states
\begin{eqnarray}
&&\vert\psi_1\rangle_{ab}=A_{1}\left[e^{-i\left(\frac{J_z}{4}+h'\right)t}\left(\cos\left(\frac{\theta}{2}\right)e^{-i\frac{J_0+h}{2}t}+\sin\left(\frac{\theta}{2}\right)e^{i\left(\frac{J_0+h}{2}t-\phi\right)}\right)^2\vert\uparrow\uparrow\rangle_{ab}\right.\nonumber\\
&&\left.+e^{-i\left(\frac{J}{2}-\frac{J_z}{4}\right)t}\left(\cos\left(\frac{\theta}{2}\right)e^{-i\frac{h}{2}t}+\sin\left(\frac{\theta}{2}\right)e^{i\left(\frac{h}{2}t-\phi\right)}\right)^2\left(\vert\uparrow\downarrow\rangle_{ab}+\vert\downarrow\uparrow\rangle_{ab}\right)\right.\nonumber\\
&&\left.+e^{-i\left(\frac{J_z}{4}-h'\right)t}\left(\cos\left(\frac{\theta}{2}\right)e^{i\frac{J_0-h}{2}t}+\sin\left(\frac{\theta}{2}\right)e^{-i\left(\frac{J_0-h}{2}t+\phi\right)}\right)^2\vert\downarrow\downarrow\rangle_{ab} \right],\label{resmeasspin121}
\end{eqnarray}
\begin{eqnarray}
&&\vert\psi_2\rangle_{ab}=A_{2}\left[e^{-i\left(\frac{J_z}{4}+h'\right)t}\left(\cos\theta+i\sin\theta\sin((h+J_0)t-\phi)\right)\vert\uparrow\uparrow\rangle_{ab}\right.\nonumber\\
&&\left.+e^{-i\left(\frac{J}{2}-\frac{J_z}{4}\right)t}\left(\cos\theta+i\sin\theta\sin(ht-\phi)\right)\left(\vert\uparrow\downarrow\rangle_{ab}+\vert\downarrow\uparrow\rangle_{ab}\right)\right.\nonumber\\
&&\left.+e^{-i\left(\frac{J_z}{4}-h'\right)t}\left(\cos\theta+i\sin\theta\sin((h-J_0)t-\phi)\right)\vert\downarrow\downarrow\rangle_{ab} \right],\label{resmeasspin122}\\
&&\vert\psi_3\rangle_{ab}=A_{3}\left[e^{-i\left(\frac{J_z}{4}+h'\right)t}\left(\cos\left(\frac{\theta}{2}\right)e^{i\frac{J_0+h}{2}t}-\sin\left(\frac{\theta}{2}\right)e^{-i\left(\frac{J_0+h}{2}t-\phi\right)}\right)^2\vert\uparrow\uparrow\rangle_{ab}\right.\nonumber\\
&&\left.+e^{-i\left(\frac{J}{2}-\frac{J_z}{4}\right)t}\left(\cos\left(\frac{\theta}{2}\right)e^{i\frac{h}{2}t}-\sin\left(\frac{\theta}{2}\right)e^{-i\left(\frac{h}{2}t-\phi\right)}\right)^2\left(\vert\uparrow\downarrow\rangle_{ab}+\vert\downarrow\uparrow\rangle_{ab}\right)\right.\nonumber\\
&&\left.+e^{-i\left(\frac{J_z}{4}-h'\right)t}\left(\cos\left(\frac{\theta}{2}\right)e^{-i\frac{J_0-h}{2}t}-\sin\left(\frac{\theta}{2}\right)e^{i\left(\frac{J_0-h}{2}t+\phi\right)}\right)^2\vert\downarrow\downarrow\rangle_{ab} \right].
\label{resmeasspin123}
\end{eqnarray}
The amplitudes of these states read
\begin{eqnarray}
&&A_1=\left[(1+\sin\theta\cos(J_0t+ht-\phi))^2+2(1+\sin\theta\cos(ht-\phi))^2\right.\nonumber\\
&&\left.+(1+\sin\theta\cos(J_0t-ht+\phi))^2\right]^{-1/2},\nonumber\\
&&A_2=\left[4\cos^2\theta+\sin^2\theta\sin^2(J_0t+ht-\phi)+2\sin^2\theta\sin^2(ht-\phi)\right.\nonumber\\
&&\left.+\sin^2\theta\sin^2(J_0t-ht+\phi)\right]^{-1/2},\nonumber\\
&&A_3=\left[(1-\sin\theta\cos(J_0t+ht-\phi))^2+2(1-\sin\theta\cos(ht-\phi))^2\right.\nonumber\\
&&\left.+(1-\sin\theta\cos(J_0t-ht+\phi))^2\right]^{-1/2}.
\label{amplitudesab}
\end{eqnarray}
Using equation (\ref{fidelity}), the fidelities of achieved states can be calculated. The results of measurements with corresponding fidelities are presented in table~\ref{tab2}.
\begin{table}[ht] 
\centering 
  \begin{tabular}{ | c| c | c | }
    \hline
State of $S_1$, $S_2$ spins & State of $S_a$, $S_b$ spins & $F$                        \\ \hline
$\vert ++\rangle_{12}$   & $\vert\psi_1\rangle_{ab}$  & $A_1^{-2}/16$           \\ \hline
$\vert +-\rangle_{12}$   & $\vert\psi_2\rangle_{ab}$  & $A_2^{-2}/16$             \\ \hline
$\vert -+\rangle_{12}$   & $\vert\psi_2\rangle_{ab}$  & $A_2^{-2}/16$             \\ \hline
$\vert --\rangle_{12}$   & $\vert\psi_3\rangle_{ab}$  & $A_3^{-2}/16$             \\ \hline
  \end{tabular}
  \caption{Results of the measurement of $S_1$, $S_2$ spins in the state (\ref{evolution12ab3}) in the direction defined by spherical angles $\theta$, $\phi$ at the moment $t$.}
  \label{tab2} 
  \end{table}

As we can see from expressions (\ref{amplitudesab}), changing angles $\theta$, $\phi$, the value of magnetic field $h$ and period of evolution allows us to control the fidelities of states prepared on the $S_a$, $S_b$ spins. For the demonstration, let us find the conditions for the preparation of the Bell states on the $S_a$ and $S_b$ spins.

\subsection{Preparation of the Bell states on the $S_a$, $S_b$ spins \label{bellstates}}

Controlling the direction of measurement of the side $S_1$, $S_2$ spins, the value of the external magnetic field, and the period of evolution, we can achieve predefined states of the $S_a$, $S_b$ spins. For example, let us find conditions that allow one to achieve the Bell states on the $S_a$, $S_b$ spins. It is easy to see from state (\ref{resmeasspin121}) that the condition $\cos\frac{\theta}{2}=-\sin\frac{\theta}{2}e^{i(ht-\phi)}$ reduces this state to the subspace spanned by $\vert\uparrow\uparrow\rangle_{ab}$, $\vert\downarrow\downarrow\rangle_{ab}$ vectors. Since the parameters $\theta$ and $\phi$ take values $\theta\in[0,\pi]$ and $\phi\in[0,2\pi]$, this equation has the following solutions
\begin{eqnarray}
\theta=\frac{\pi}{2},\quad ht-\phi=\pi.
\label{parametersab}
\end{eqnarray}
Then states (\ref{resmeasspin121}), (\ref{resmeasspin122}) and (\ref{resmeasspin123}) modulo a global phase take the form
\begin{eqnarray}
&&\vert\psi_1\rangle_{ab}=\frac{1}{\sqrt{2}}\left[\vert\uparrow\uparrow\rangle_{ab} + e^{2ih't}\vert\downarrow\downarrow\rangle_{ab} \right],\label{resmeasspin1211}\\
&&\vert\psi_2\rangle_{ab}=\frac{1}{\sqrt{2}}\left[\vert\uparrow\uparrow\rangle_{ab}+e^{i\left(2h't+\pi\right)}\vert\downarrow\downarrow\rangle_{ab} \right],\label{resmeasspin1222}\\
&&\vert\psi_3\rangle_{ab}=\frac{1}{\sqrt{2}\sqrt{\cos^4\left(\frac{J_0t}{2}\right)+1}}\left[e^{-i\left(\frac{J_z}{4}+h'\right)t}\cos^2\left(\frac{J_0t}{2}\right)\left(\vert\uparrow\uparrow\rangle_{ab}+e^{2ih't}\vert\downarrow\downarrow\rangle_{ab}\right)\right.\nonumber\\
&&\left.+e^{-i\left(\frac{J}{2}-\frac{J_z}{4}\right)t}\left(\vert\uparrow\downarrow\rangle_{ab}+\vert\downarrow\uparrow\rangle_{ab}\right) \right].
\label{resmeasspin1233}
\end{eqnarray}
States (\ref{resmeasspin1211}), (\ref{resmeasspin1222}) are maximally entangled and become Bell states $\vert\Phi^{\pm}\rangle$ for $h't=\frac{\pi}{2}n$, where $n\in\mathds{Z}$.
Entanglement of state (\ref{resmeasspin1233}) depends on time and coupling constant between spins as follows
\begin{eqnarray}
C\left(\vert\psi_3\rangle_{ab}\right)=\frac{1}{1+\cos^4\left(\frac{J_0t}{2}\right)}\left[1+\cos^8\left(\frac{J_0t}{2}\right)-2\cos^4\left(\frac{J_0t}{2}\right)\cos(J_z-J)t\right]^{1/2}.
\label{concurensepsi3ab}
\end{eqnarray}
\begin{figure}[!!h]
\includegraphics[scale=1.00, angle=0.0, clip]{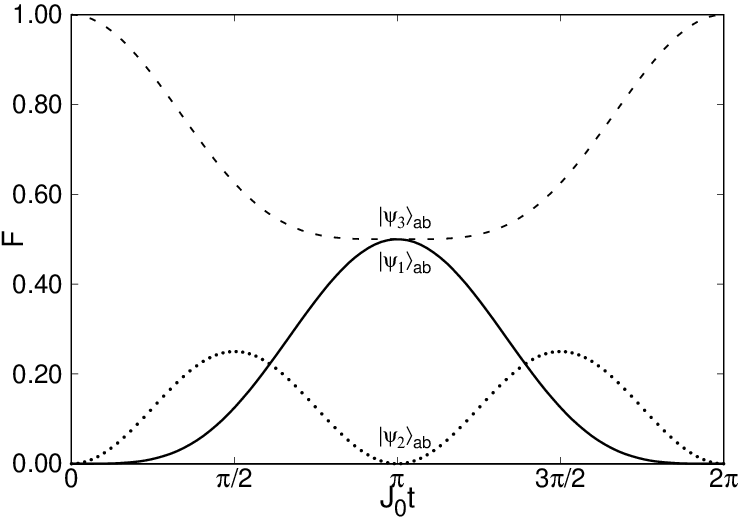}
\caption{Time dependencies of the fidelities to obtaining each of states (\ref{resmeasspin1211}), (\ref{resmeasspin1222}), (\ref{resmeasspin1233}) after measuring the side spins.}
\label{fidelityab}
\end{figure}
In the case of $J_0t=\pi+2\pi n$, modulo a global phase this state becomes the $\vert\Psi^+\rangle$ Bell state. Based on the definitions of amplitudes (\ref{amplitudesab}) and equations from table~\ref{tab2},
we obtain the fidelities of each of states (\ref{resmeasspin1211}), (\ref{resmeasspin1222}), (\ref{resmeasspin1233}) after measuring of the side spins. Substituting parameters (\ref{parametersab}) in these equations
we obtain
\begin{eqnarray}
&&F\left(\vert\psi_1\rangle_{ab}\right)=\frac{1}{2}\sin^4\left(\frac{J_0t}{2}\right),\nonumber\\
&&F\left(\vert\psi_2\rangle_{ab}\right)=\frac{1}{4}\sin^2\left(J_0t\right),\nonumber\\
&&F\left(\vert\psi_3\rangle_{ab}\right)=\frac{1}{2}\left(\cos^4\left(\frac{J_0t}{2}\right)+1\right).
\label{fidelities2}
\end{eqnarray}
These dependencies change with the period $2\pi$ with respect to the parameter $J_0t$ (see, Fig.~\ref{fidelityab}). As we can see, at the moment $T=\pi/J_0$ with fidelity $0.5$, we obtain both $\vert\psi_1\rangle_{ab}$ (\ref{resmeasspin1211}) and $\vert\Psi^+\rangle$ states. It also follows from expression (\ref{parametersab}) that the $S_1$, $S_2$ spins should be measured in the direction defined by the spherical angles $\theta=\pi/2$, $\phi=(h/J_0-1)\pi$. In addition, in order to achieve the $\vert\Phi^+\rangle$ and $\vert\Phi^-\rangle$ Bell states from the state (\ref{resmeasspin1211}), the magnetic field of the values $h'=0$ and $J_0/2$ should be applied to the $S_a$, $S_b$ spins. The $\vert\Phi^{\pm}\rangle$ Bell state can be obtained from the state (\ref{resmeasspin1222}) with fidelity $0.25$ if we make the measurements of side spins in the moments $T=\pi/(2J_0)$ and $3\pi/(2J_0)$. In this case, the side spins should be measured in the directions defined by the sets of parameters: 1. $\theta=\pi/2$, $\phi=\left(h/(2J_0)-1\right)\pi$ for $T=\pi/(2J_0)$ and 2. $\theta=\pi/2$, $\phi=\left(3h/(2J_0)-1\right)\pi$ for $T=3\pi/(2J_0)$. Then the magnetic field $h'$ should be given the values: 1. $h'=0$ for $\vert\Phi^-\rangle $ Bell state and 2. $h'=J_0$ for $\vert\Phi^+\rangle$ Bell state. Finally, it is worth noting that a similar situation exists when we put the following conditions $\theta=\pi/2$ and $ht-\phi=\pi$ on measurement direction. In this case, the state $\vert\psi_1\rangle_{ab}$ takes the form defined by expression (\ref{resmeasspin1233}) and vice versa.

\begin{figure}[!!h]
\centerline{\includegraphics[scale=0.89, angle=0.0, clip]{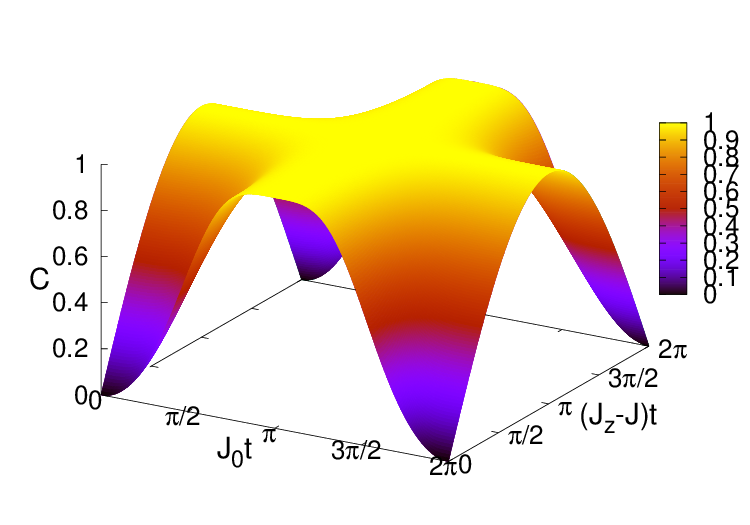}}
\centerline{\includegraphics[scale=0.89, angle=0.0, clip]{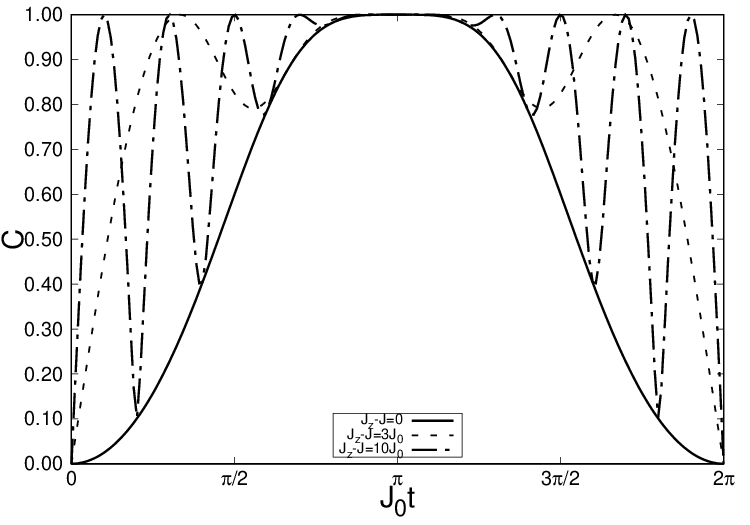}}
\caption{Time dependence of concurrence (\ref{concurensepsi3ab}) of state $\vert\psi_3\rangle$ on the ratio between the interaction parameters $J_0$ and $J_z-J$.
Increasing the anisotropy and the value of interaction between the $S_a$ and $S_b$ spins allows us to achieve maximally entangled states faster. In the down frame, we present the behavior of concurrence
for different ratios of interaction parameters.}
\label{entabdyb12}
\end{figure}

Finally, we depict the time dependence of concurrence (\ref{concurensepsi3ab}) of state $\vert\psi_3\rangle$ on the ratio between the interaction parameters $J_0$ and $J_z-J$ (Fig.~\ref{entabdyb12}).
It is easy to see that the stronger the anisotropy and the value of interaction between the $S_a$ and $S_b$ spins, the faster the state $\vert\psi_3\rangle$ becomes maximally entangled.

\section{Preparation of entanglement states on the $S_1$, $S_2$ spins \label{statesspin12}}

In this section, we examine the states prepared on the $S_1$, $S_2$ spins depending on the measurement direction of the $S_a$, $S_b$ spins. For this purpose, we rewrite state (\ref{evolution12ab2}) in the following way
\begin{eqnarray}
\vert\psi(t)\rangle=\frac{1}{2}\left(\vert\phi_1\rangle_{12}\vert\uparrow\uparrow\rangle_{ab}+ \vert\phi_2\rangle_{12}\left(\vert\uparrow\downarrow\rangle_{ab}+\vert\downarrow\uparrow\rangle_{ab}\right) + \vert\phi_3\rangle_{12}\vert\downarrow\downarrow\rangle_{ab}\right),
\label{initstate12abevolut}
\end{eqnarray}
where we introduce the following notations
{\small
\begin{eqnarray}
&&\vert\phi_1\rangle_{12}=\frac{1}{2}\left[e^{-i(\frac{J_z}{4}+J_0+h+h')t}\vert\uparrow\uparrow\rangle_{12}+e^{-i(\frac{J_z}{4}+h')t}(\vert\uparrow\downarrow\rangle_{12}+\vert\downarrow\uparrow\rangle_{12})+e^{-i(\frac{J_z}{4}-J_0-h+h')t}\vert\downarrow\downarrow\rangle_{12}     \right],\nonumber\\
&&\vert\phi_2\rangle_{12}=\frac{1}{2}\left[e^{-i(\frac{J}{2}-\frac{J_z}{4}+h)t}\vert\uparrow\uparrow\rangle_{12}+e^{-i(\frac{J}{2}-\frac{J_z}{4})t}(\vert\uparrow\downarrow\rangle_{12}+\vert\downarrow\uparrow\rangle_{12})+e^{-i(\frac{J}{2}-\frac{J_z}{4}-h)t}\vert\downarrow\downarrow\rangle_{12}     \right],\nonumber\\
&&\vert\phi_3\rangle_{12}=\frac{1}{2}\left[e^{-i(\frac{J_z}{4}-J_0+h-h')t}\vert\uparrow\uparrow\rangle_{12}+e^{-i(\frac{J_z}{4}-h')t}(\vert\uparrow\downarrow\rangle_{12}+\vert\downarrow\uparrow\rangle_{12})+e^{-i(\frac{J_z}{4}+J_0-h-h')t}\vert\downarrow\downarrow\rangle_{12}     \right].\nonumber\\
\label{purestateof12spins}
\end{eqnarray}}
Measuring the spins $S_a$, $S_b$ on the basis $\vert\uparrow\uparrow\rangle_{ab}$, $\vert\uparrow\downarrow\rangle_{ab}$, $\vert\downarrow\uparrow\rangle_{ab}$ $\vert\downarrow\downarrow\rangle_{ab}$ with fidelities $F\left(\vert\phi_1\rangle_{12}\right)=1/4$, $F\left(\vert\phi_2\rangle_{12}\right)=1/2$ and $F\left(\vert\phi_3\rangle_{12}\right)=1/4$ we obtain states $\vert\phi_1\rangle_{12}$, $\vert\phi_2\rangle_{12}$ and $\vert\phi_3\rangle_{12}$ of the $S_1$, $S_2$ spins, respectively. The value of entanglement (\ref{wootterspure}) of each states is equal to $C=0$. However, making measurements of the $S_a$, $S_b$ spins on another basis states allow one to achieve entangled states of the $S_1$, $S_2$ spins. Using relations between basis states $\vert\uparrow\rangle$, $\vert\downarrow\rangle$ and $\vert +\rangle$, $\vert -\rangle$ (\ref{newbasisstate}) for $S_a$, $S_b$ spins, state (\ref{initstate12abevolut}) takes the form (\ref{appa11}) (see, Appendix~\ref{appa}). Measuring $S_a$, $S_b$ spins, we obtain that the achieved states of the $S_1$, $S_2$ spins addition depend on the difference between interaction couplings $J-J_z$. For example, in the case of isotropic interaction between $S_a$ and $S_b$ spins ($J=J_z$), the achieved states take the form similar to states (\ref{resmeasspin121}), (\ref{resmeasspin122}) and (\ref{resmeasspin123}) with replacement $h$ on $h'$ and vice versa. Here, the Bell states are prepared in the same way described in the previous section.

\section{Conclusions \label{conc}}

We have considered the preparation of entangled pure quantum states on the Ising-Heisenberg diamond spin cluster placed in the magnetic field. This cluster consists of two central spins described by the anisotropic Heisenberg interaction
which interacts that two side spins via the Ising model. It is worth noting that the ions in the copper-based compounds mentioned in the introduction
are arranged in a spin-1/2 diamond chain. The interaction between each spin in the chain is described by the Heisenberg model. However, to simplify calculations, we have considered a simpler model where the side spins interact
with central spins via the Ising model. Depending on the initial state, we have studied the evolution of this system. Namely, we have examined the preparation of entangled states on the central spins when the side spins
are measured and vice versa. Due to the fact that parts of Hamiltonian which describe the central and side spins mutually commute, we can independently study the evolution of one subsystem without intertwining it with another subsystem.
We have considered such evolution when side spins are in the stationary state. The influence of the side spins on the evolution of the central spins is similar to the presence of an effective magnetic field. We have obtained
the conditions to achieve the entangled states on the central spins.

In the case when the whole system evolves, we have shown that the direction, in which the spins of one subsystem are measured, affects the form and entanglement of the achieved states on the other subsystem.
Firstly, we have investigated the preparation of pure entangled states on the central spins depending on the measurement direction of the side spins. The fidelities of these states as a function of the period of evolution,
parameters of Hamiltonian and the measurement direction of the side spins have been calculated. For example, we have obtained conditions and fidelities for the preparation of the $\vert\Phi^{\pm}\rangle$ and $\vert\Psi^+\rangle$
Bell states. We have also examined the preparation of the states on side spins depending on the measurement direction of the central spins. It has been shown that the entanglement of states achieved on side spins depends
on the measurement direction of the central spins. There is a measurement direction in which all achieved states are separated and another measurement direction that allows one to prepare the maximally entangled states.

\section{Acknowledgements}

This work was supported by Project 77/02.2020 from National Research Foundation of Ukraine.

\begin{appendices}

\section{Eigenstates and eigenvalues of the diamond spin cluster \label{appegen}}
\setcounter{equation}{0}
\renewcommand{\theequation}{A\arabic{equation}}

The fact that Hamiltonians (\ref{hamab}), (\ref{ham12}) and (\ref{hamint}) mutually commute, we can easily obtain the eigenstates and corresponding eigenvalues of Hamiltonian (\ref{hamiltonian}).
These eigenstates and eigenvalues have the following form
\begin{align}
&\vert\psi_1\rangle = \vert\uparrow\uparrow\rangle_{{\tiny 12}}\vert\uparrow\uparrow\rangle_{\small{ab}}, && E_{1}=h+\frac{J_z}{4}+h'+J_0,\nonumber\\
&\vert\psi_2\rangle = \vert\uparrow\uparrow\rangle_{\small{12}}\frac{1}{\sqrt{2}}\left(\vert\uparrow\downarrow\rangle+\vert\downarrow\uparrow\rangle\right)_{ab}, && E_{2}=h+\frac{J}{2}-\frac{J_z}{4},\nonumber\\
&\vert\psi_3\rangle = \vert\uparrow\uparrow\rangle_{\small{12}}\frac{1}{\sqrt{2}}\left(\vert\uparrow\downarrow\rangle-\vert\downarrow\uparrow\rangle\right)_{ab}, && E_{3}=h-\frac{J}{2}-\frac{J_z}{4},\nonumber\\
&\vert\psi_4\rangle = \vert\uparrow\uparrow\rangle_{{\tiny 12}}\vert\downarrow\downarrow\rangle_{\small{ab}}, && E_{4}=h+\frac{J_z}{4}-h'-J_0,\nonumber\\
&\vert\psi_5\rangle = \vert\uparrow\downarrow\rangle_{{\tiny 12}}\vert\uparrow\uparrow\rangle_{\small{ab}}, && E_{5}=\frac{J_z}{4}+h',\nonumber\\
&\vert\psi_6\rangle = \vert\uparrow\downarrow\rangle_{\small{12}}\frac{1}{\sqrt{2}}\left(\vert\uparrow\downarrow\rangle+\vert\downarrow\uparrow\rangle\right)_{ab}, && E_{6}=\frac{J}{2}-\frac{J_z}{4},\nonumber\\
&\vert\psi_7\rangle = \vert\uparrow\downarrow\rangle_{\small{12}}\frac{1}{\sqrt{2}}\left(\vert\uparrow\downarrow\rangle-\vert\downarrow\uparrow\rangle\right)_{ab}, && E_{7}=-\frac{J}{2}-\frac{J_z}{4},\nonumber\\
&\vert\psi_8\rangle = \vert\uparrow\downarrow\rangle_{{\tiny 12}}\vert\downarrow\downarrow\rangle_{\small{ab}}, && E_{8}=\frac{J_z}{4}-h',\nonumber\\
&\vert\psi_9\rangle = \vert\downarrow\uparrow\rangle_{{\tiny 12}}\vert\uparrow\uparrow\rangle_{\small{ab}}, && E_{9}=\frac{J_z}{4}+h',\nonumber\\
&\vert\psi_{10}\rangle = \vert\downarrow\uparrow\rangle_{\small{12}}\frac{1}{\sqrt{2}}\left(\vert\uparrow\downarrow\rangle+\vert\downarrow\uparrow\rangle\right)_{ab}, && E_{10}=\frac{J}{2}-\frac{J_z}{4},\nonumber\\
&\vert\psi_{11}\rangle = \vert\downarrow\uparrow\rangle_{\small{12}}\frac{1}{\sqrt{2}}\left(\vert\uparrow\downarrow\rangle-\vert\downarrow\uparrow\rangle\right)_{ab}, && E_{11}=-\frac{J}{2}-\frac{J_z}{4},\nonumber\\
&\vert\psi_{12}\rangle = \vert\downarrow\uparrow\rangle_{{\tiny 12}}\vert\downarrow\downarrow\rangle_{\small{ab}}, && E_{12}=\frac{J_z}{4}-h',\nonumber\\
&\vert\psi_{13}\rangle = \vert\downarrow\downarrow\rangle_{{\tiny 12}}\vert\uparrow\uparrow\rangle_{\small{ab}}, && E_{13}=-h+\frac{J_z}{4}+h'-J_0,\nonumber\\
&\vert\psi_{14}\rangle = \vert\downarrow\downarrow\rangle_{\small{12}}\frac{1}{\sqrt{2}}\left(\vert\uparrow\downarrow\rangle+\vert\downarrow\uparrow\rangle\right)_{ab}, && E_{14}=-h+\frac{J}{2}-\frac{J_z}{4},\nonumber\\
&\vert\psi_{15}\rangle = \vert\downarrow\downarrow\rangle_{\small{12}}\frac{1}{\sqrt{2}}\left(\vert\uparrow\downarrow\rangle-\vert\downarrow\uparrow\rangle\right)_{ab}, && E_{15}=-h-\frac{J}{2}-\frac{J_z}{4},\nonumber\\
&\vert\psi_{16}\rangle = \vert\downarrow\downarrow\rangle_{{\tiny 12}}\vert\downarrow\downarrow\rangle_{\small{ab}}, && E_{16}=-h+\frac{J_z}{4}-h'+J_0.
\label{eigenvaleigenstate}
\end{align}
The states of subsystems are indicated by the subscripts. The states of $S_1$, $S_2$ and $S_a$, $S_b$ spins are denoted by the subscripts $12$ and $ab$, respectively.

\section{Relations between $\vert\uparrow\rangle$, $\vert\downarrow\rangle$ and $\vert +\rangle$, $\vert -\rangle$ basis states of two spins \label{appproject}}
\setcounter{equation}{0}
\renewcommand{\theequation}{B\arabic{equation}}

In this appendix, we present the relations between different basis states of two spins. Equations (\ref{possnegstates}) determine the states of spin-$1/2$ projected in the positive and negative direction of the axis defined by the spherical angles $\theta$ and $\phi$. The inverse relations to states (\ref{possnegstates}) have the form
\begin{eqnarray}
\vert\uparrow\rangle=\cos\left(\frac{\theta}{2}\right)\vert +\rangle-\sin\left(\frac{\theta}{2}\right)e^{i\phi}\vert -\rangle,\quad \vert\downarrow\rangle=\sin\left(\frac{\theta}{2}\right)e^{-i\phi}\vert +\rangle+\cos\left(\frac{\theta}{2}\right)\vert -\rangle.
\end{eqnarray}
Using these relations the basis states of two $S_1$, $S_2$ spins can be rewritten as follows
\begin{eqnarray}
&&\vert\uparrow\uparrow\rangle_{12}=\cos^2\left(\frac{\theta}{2}\right)\vert ++\rangle_{12} -\cos\left(\frac{\theta}{2}\right)\sin\left(\frac{\theta}{2}\right)e^{i\phi}\left(\vert+-\rangle_{12}+\vert -+\rangle_{12}\right)\nonumber\\
&&+\sin^2\left(\frac{\theta}{2}\right)e^{2i\phi}\vert -- \rangle_{12},\nonumber\\
&&\vert\uparrow\downarrow\rangle_{12}=\cos\left(\frac{\theta}{2}\right)\sin\left(\frac{\theta}{2}\right)e^{-i\phi}\vert ++\rangle_{12} +\cos^2\left(\frac{\theta}{2}\right)\vert+-\rangle_{12}\nonumber\\
&&-\sin^2\left(\frac{\theta}{2}\right)\vert -+\rangle_{12}-\cos\left(\frac{\theta}{2}\right)\sin\left(\frac{\theta}{2}\right)e^{i\phi}\vert -- \rangle_{12},\nonumber\\
&&\vert\downarrow\uparrow\rangle_{12}=\cos\left(\frac{\theta}{2}\right)\sin\left(\frac{\theta}{2}\right)e^{-i\phi}\vert ++\rangle_{12} -\sin^2\left(\frac{\theta}{2}\right)\vert+-\rangle_{12}\nonumber\\
&&+\cos^2\left(\frac{\theta}{2}\right)\vert -+\rangle_{12}-\cos\left(\frac{\theta}{2}\right)\sin\left(\frac{\theta}{2}\right)e^{i\phi}\vert -- \rangle_{12},\nonumber\\
&&\vert\downarrow\downarrow\rangle_{12}=\sin^2\left(\frac{\theta}{2}\right)e^{-2i\phi}\vert ++\rangle_{12} +\cos\left(\frac{\theta}{2}\right)\sin\left(\frac{\theta}{2}\right)e^{-i\phi}\left(\vert+-\rangle_{12}+\vert -+\rangle_{12}\right)\nonumber\\
&&+\cos^2\left(\frac{\theta}{2}\right)\vert -- \rangle_{12}.
\label{newbasisstate}
\end{eqnarray}

\section{State of the system in the basis $\vert +\rangle$, $\vert -\rangle$ for $S_a$ and $S_b$ spins \label{appa}}
\setcounter{equation}{0}
\renewcommand{\theequation}{C\arabic{equation}}

In this appendix, using relations between basis states $\vert\uparrow\rangle$, $\vert\downarrow\rangle$ and $\vert +\rangle$, $\vert -\rangle$ (\ref{newbasisstate}) for $S_a$, $S_b$ spins, we
rewrite state (\ref{initstate12abevolut}) in the form
{\footnotesize
\begin{eqnarray}
&&\vert\psi(t)\rangle=\frac{1}{4}e^{-i\left(\frac{J_z}{4}+h'\right)t}\nonumber\\
&&\times\left[ e^{-iht}\left(\cos^2\left(\frac{\theta}{2}\right)e^{-iJ_0t} + \sin\left(\theta\right)e^{-i(\phi-h't)}e^{-i\left(\frac{J}{2}-\frac{J_z}{2}\right)t} + \sin^2\left(\frac{\theta}{2}\right)e^{-2i(\phi-h't)}e^{iJ_0t}  \right)   \vert\uparrow\uparrow\rangle_{12} \right.\nonumber\\
&&\left.+\left(\cos^2\left(\frac{\theta}{2}\right) + \sin\left(\theta\right)e^{-i(\phi-h't)}e^{-i\left(\frac{J}{2}-\frac{J_z}{2}\right)t} + \sin^2\left(\frac{\theta}{2}\right)e^{-2i(\phi-h't)}  \right)\left( \vert\uparrow\downarrow\rangle_{12}+\vert\downarrow\uparrow\rangle_{12}\right)\right.\nonumber\\
&&\left.+e^{iht}\left(\cos^2\left(\frac{\theta}{2}\right)e^{iJ_0t} + \sin\left(\theta\right)e^{-i(\phi-h't)}e^{-i\left(\frac{J}{2}-\frac{J_z}{2}\right)t} + \sin^2\left(\frac{\theta}{2}\right)e^{-2i(\phi-h't)}e^{-iJ_0t}  \right)   \vert\downarrow\downarrow\rangle_{12} \right]\vert ++\rangle_{ab}\nonumber\\
&&+\frac{1}{4}e^{-i\frac{J_z}{4}t}\left[ e^{-iht}\left(\cos\theta e^{-i\left(\frac{J}{2}-\frac{J_z}{2}\right)t} + i \sin\theta\sin\left((h'+J_0)t-\phi\right) \right)   \vert\uparrow\uparrow\rangle_{12} \right.\nonumber\\
&&\left.+\left(\cos\theta e^{-i\left(\frac{J}{2}-\frac{J_z}{2}\right)t} + i \sin\theta\sin\left(h't-\phi\right) \right)\left( \vert\uparrow\downarrow\rangle_{12}+\vert\downarrow\uparrow\rangle_{12}\right) \right.\nonumber\\
&&\left.+e^{iht}\left(\cos\theta e^{-i\left(\frac{J}{2}-\frac{J_z}{2}\right)t} + i \sin\theta\sin\left((h'-J_0)t-\phi\right) \right)   \vert\downarrow\downarrow\rangle_{12} \right]\left(\vert +-\rangle_{ab}+\vert -+\rangle_{ab}\right)\nonumber\\
&&+\frac{1}{4}e^{-i\left(\frac{J_z}{4}-h'\right)t}\nonumber\\
&&\times\left[ e^{-iht}\left(\sin^2\left(\frac{\theta}{2}\right)e^{2i(\phi-h't)}e^{-iJ_0t} - \sin\left(\theta\right)e^{i(\phi-h't)}e^{-i\left(\frac{J}{2}-\frac{J_z}{2}\right)t} + \cos^2\left(\frac{\theta}{2}\right)e^{iJ_0t}  \right)   \vert\uparrow\uparrow\rangle_{12} \right.\nonumber\\
&&\left.+\left(\sin^2\left(\frac{\theta}{2}\right)e^{2i(\phi-h't)} - \sin\left(\theta\right)e^{i(\phi-h't)}e^{-i\left(\frac{J}{2}-\frac{J_z}{2}\right)t} + \cos^2\left(\frac{\theta}{2}\right)  \right)\left( \vert\uparrow\downarrow\rangle_{12}+\vert\downarrow\uparrow\rangle_{12}\right)\right.\nonumber\\
&&\left.+e^{iht}\left(\sin^2\left(\frac{\theta}{2}\right)e^{2i(\phi-h't)}e^{iJ_0t} - \sin\left(\theta\right)e^{i(\phi-h't)}e^{-i\left(\frac{J}{2}-\frac{J_z}{2}\right)t} + \cos^2\left(\frac{\theta}{2}\right)e^{-iJ_0t}  \right)   \vert\downarrow\downarrow\rangle_{12} \right]\vert --\rangle_{ab}.\nonumber\\
\label{appa11}
\end{eqnarray}
}

\end{appendices}

{}

\end{document}